\documentclass{article}
\usepackage{spconf,amsmath,graphicx,hyperref}
\usepackage{cite}
\usepackage{url}
\usepackage{amssymb,amsfonts}
\usepackage{algorithmic}
\usepackage{textcomp}
\usepackage{subcaption}
\usepackage{xcolor}
\usepackage{tikz,pgfplots}

\usepackage{booktabs}
\title{SECOND-ORDER OPTIMIZATION OF VARIABLE PROJECTION SVM MODELS AND ROAD ABNORMALITY DETECTION}
\name{Andrea Angino$^{\star}$ \qquad Matthias Voigt$^{\star}$ \qquad 
      Rolf Krause$^{\dagger,\star}$ \qquad Tam\'as D\'ozsa$^{\star, \ddagger}$}

\address{$^{\star}$ Faculty of Mathematics and Computer Science, UniDistance Suisse, Brig, Switzerland \\
         $^{\dagger}$ CEMSE Division,  KAUST, Thuwal, Saudi Arabia \\
         $^{\ddagger}$ E\"otv\"os Lor\'and University, Faculty of Informatics, Department of Numerical Analysis
         }
\begin{document}
%\ninept
\onecolumn\begin{titlepage}
\thispagestyle{empty}

\begin{center}

\vspace*{2.5cm}

{\LARGE\bfseries
Second-order optimization of variable projection SVM models\\
and road abnormality detection
\par}

\vspace{1.2cm}

{\large
Andrea Angino, Matthias Voigt, Rolf Krause, Tam\'as D\'ozsa
\par}

\vspace{2cm}

{\large\bfseries Author Accepted Manuscript}

\vspace{0.7cm}

\begin{minipage}{0.82\textwidth}
\centering
This document is the accepted version of the manuscript and is not the
final published version.

\vspace{0.8cm}

The final published article is available at:

\medskip

\href{https://doi.org/10.1109/ICASSP55912.2026.11463454}
{\texttt{https://doi.org/10.1109/ICASSP55912.2026.11463454}.}

\vspace{0.8cm}

Please cite the final published version when referring to this work: \medskip \\
A. Angino, M. Voigt, R. Krause, and T. D\'ozsa. Second-order optimization of variable projection
SVM models and road abnormality detection. In \textit{ICASSP 2026 – 2026 IEEE International
Conference on Acoustics, Speech and Signal Processing (ICASSP)}, pages 1641–1645, Barcelona,
Spain, 2026.

\vspace{1.2cm}

\textcopyright{} 2026 IEEE.
\end{minipage}

\vfill

\end{center}
\end{titlepage}
\clearpage

\twocolumn

\maketitle

\begin{abstract}
We introduce a novel second-order optimization framework for minimizing so-called variable projection functionals. 
%These problems have traditionally been treated using first-order schemes such as gradient descent and its stochastic variants.
We demonstrate that the proposed framework is especially useful for the training of variable projection based kernel methods. In particular, the problem of efficiently training variable projection support vector machines (VP-SVMs) is considered. 
%VP-SVM is a recently introduced machine learning model capable of solving a variety of classification problems in an interpretable manner, making it an ideal algorithm for several safety-critical signal processing applications.
We show the effectiveness of the proposed training methodology in a real-world application, namely we demonstrate how second-order trust region algorithms can be used to train VP-SVM models to recognize road surface abnormalities based on 1D signals obtained from a tire sensor.
%We analyze the performance of the proposed methods using this dataset and conclude that they vastly outperform traditional counterparts. 
\end{abstract}
\section{Introduction}
\label{sec:intro}

Artificial intelligence (AI) methods have become a crucial component of many state-of-the-art signal processing algorithms in recent years. The main advantage of ML methods is their ability to approximate nonlinear operators in a purely data-driven manner. Unfortunately, ML methods, especially those that rely on deep learning, often suffer from a lack of explainability. This is mainly due to the size of current models: Deep neural networks often contain tens of millions of parameters~\cite{lipton2018mythos}. The lack of interpretable parameters makes the deployment of ML models difficult in safety-critical applications such as fault detection in critical infrastructure. 

% In order to remedy the above problems, the field of explainable artificial intelligence (XAI) has recently enjoyed growing attention. One exciting direction of XAI is the field of transparent AI methods~\cite{lipton2018mythos} which aim to construct novel ML architectures whose parameters can be directly related to interpretable quantities. Recent examples of transparent methods include physics-informed models~\cite{karniadakis2021physics}, wavelet feature based approaches~\cite{karniadakis2021physics} and variable projection (VP) based ML schemes~\cite{KOVACS, dozsa2024variable, weightedHermdozsa}. 

Although many papers~\cite{lipton2018mythos,karniadakis2021physics, KOVACS,dozsa2024variable,weightedHermdozsa} focus on introducing new transparent AI architectures to mitigate these problems, training methods that exploit the structural properties associated with them are less frequent. In this work, we demonstrate that second-order trust-region (TR) methods can be an effective tool to optimize the parameters of so-called variable projection support vector machines (VP-SVMs), see~\cite{dozsa2024variable}. VP-SVMs were recently introduced in~\cite{dozsa2024variable} to improve the interpretability and performance of classical SVM classifiers. VP-SVMs contain interpretable parameters and therefore fit well with the transparent AI paradigm. The key component of a VP-SVM model is an adaptive feature transformation step, which transforms the data samples to be classified to a low-dimensional space. The (physically meaningful) parameters of this transformation are optimized together with the weights of the SVM classifier. %VP-SVM may be used together with Mercer kernels (see~\cite{chapelle2007training}), resulting in SVM models, where the kernel matrix also depends on learnable parameters.

SVMs (and kernel methods in general) are difficult to train on large datasets. Indeed, the associated kernel matrix scales with the number of training signals~\cite{chapelle2007training, dozsa2024variable}. In the case of VP-SVM, this problem is even more significant since the kernel also depends on learnable parameters. In order to make the training problem numerically tractable for VP-SVM, in~\cite{dozsa2024variable}, a reduced objective is proposed together with the use of stochastic subgradient methods. This, however, may lead to suboptimal classification and a lack of interpretability. 

In this paper, we propose second-order training methods to address these challenges. In particular, we show that the training time of a VP-SVM model that can recognize road surface abnormalities (such as potholes or bumps), see \cite {roadbnorm, roadabnorm2}, can be greatly reduced with a properly chosen second-order training scheme. In fact, as our experiments show, the generalization abilities of the classifier will improve significantly, despite the reduction to the cost of training. 

In particular, we derive analytic formulas for the second-order derivatives of Moore-Penrose pseudo-inverses and variable projection operators as well as Hessians of primal quadratic VP-SVM objectives. Then, we show how second-order optimization schemes can be used to efficiently train VP-SVM classifiers to recognize road surface abnormalities. For reproducibility, we refer to our implementation~\cite{code} that also contains comparisons between different training schemes. Although our results are stated for VP-SVM, they can be extended to any kernel-based models, where the kernels depend on learnable parameters.

%The automatic detection of obstacles (such as potholes or bumps) on the road surface is an important problem for improving the safety conditions and reducing the maintenance costs of automotive travel~\cite{roadbnorm}. 
%Studies~\cite {roadbnorm, roadabnorm2} show that variable projection based ML algorithms can effectively be used for this purpose. 
%In this work, we show that VP-SVM models, which match the performance of previous methods~\cite{roadbnorm}, can be trained in an efficient manner using second-order training schemes.
\iffalse
The main contributions of this paper are:
\begin{enumerate}
    \item We derive analytic formulas for the second-order derivatives of Moore-Penrose pseudo-inverses and variable projection operators as well as Hessians of primal quadratic VP-SVM objectives. 
    \item We show how second-order optimization schemes can be used to efficiently train VP-SVM classifiers to recognize road surface abnormalities. %Model performance matches the state-of-the-art described in~\cite{roadbnorm}, but the use of the VP-SVM architecture improves interpretability. 
    For reproducibility, we refer to our implementation~\cite{code} that also contains comparisons between different training schemes. 
    %\item We provide publicly available implementations of our training algorithms to ensure reproducibility.
\end{enumerate}
\fi
%It is important to note that, although our results are stated for a specific classification scheme (VP-SVM), they can be useful for the training of any kernel based models, where the kernels depend on learnable parameters.

\section{Variable projection operators}

First, we consider separable nonlinear least squares (SNLLS) problems, VP operators, and their connection to feature extraction in ML. Let $\mathcal{H}$ be a Hilbert space of functions $X \to \mathbb{R}$ with $X \subseteq \mathbb{R}$. Suppose $\{\phi_{\boldsymbol{\eta}}^1,\ldots, \phi_{\boldsymbol{\eta}}^m\}$ is a linearly independent system in $\mathcal{H}$ for all parameter vectors $\boldsymbol{\eta} \in \mathbb{R}^p$. Henceforth, we shall consider the vectors 
$\varphi_k(\boldsymbol{\eta}) := {[\phi_{\boldsymbol{\eta}}^k(x_\ell)]}_{\ell=1}^n \in \mathbb{R}^n$
%$\varphi_k(\boldsymbol{\eta})_j := \phi_{\boldsymbol{\eta}, k}(x_\ell)$, 
where $x_\ell \in X$ for $\ell=1,\ldots,n$ with $x_1 < \ldots < x_n$. We shall assume that the sampling points $x_\ell$ are chosen such that the vectors $    {\{\varphi_k(\boldsymbol{\eta})\}}_{k=1}^m$ remain linearly independent in $\mathbb{R}^n$.

%Let $(\mathcal{H}, \langle \cdot, \cdot \rangle)$ be an arbitrary Hilbert space and let $\boldsymbol{f} \in \mathcal{H}$. Although the theory reviewed here holds for general $\mathcal{H}$, for the sake of simplicity we henceforth assume $\mathcal{H} := (\mathbb{R}^n, \langle \cdot, \cdot \rangle)$, where $\langle \cdot, \cdot \rangle$ denotes the usual Euclidean inner product. Furthermore, let $\varphi_1(\boldsymbol{\eta}), \varphi_2(\boldsymbol{\eta}), \ldots, \varphi_m(\boldsymbol{\eta}) \in \mathbb{R}^n\ (m \in \mathbb{N}, m \ll n)$ be a linearly independent system of function values of twice continuously differentiable functions $\varphi_1,\ldots,\varphi_m: \mathbb{R}^p \to \mathbb{R}^n$. %with respect to the parameter $\boldsymbol{\eta} \in \mathbb{R}^p \ (p \in \mathbb{N})$.

Since $\varphi_k(\boldsymbol{\eta}), k=1,\ldots,m$ are linearly independent, they span an $m$-dimensional subspace $G(\boldsymbol{\eta}) := \operatorname{span} \{ \varphi_k(\boldsymbol{\eta}) \;: \; k=1,\ldots,m \} \subset \mathbb{R}^n$. It is well known (see, e.g.,~\cite{petrusev}) that for a fixed $\boldsymbol{\eta}$ there exists $\widehat{\boldsymbol{f}} \in G(\boldsymbol{\eta})$, such that ${\|\boldsymbol{f} - \widehat{\boldsymbol{f}} \|}_2$ is minimal. In fact, with the standard Euclidean inner product $\langle \cdot,\cdot \rangle$, $\langle \boldsymbol{f} - \widehat{\boldsymbol{f}}, \boldsymbol{g} \rangle = 0$ ($\boldsymbol{g} \in G(\boldsymbol{\eta})$) holds, that is, $\widehat{\boldsymbol{f}}$ is the orthogonal projection of $\boldsymbol{f}$ onto $G(\boldsymbol{\eta})$.  The projection operator $P_{\boldsymbol{\eta}} : \mathbb{R}^n \to \mathbb{R}^n$ which satisfies $P_{\boldsymbol{\eta}} \boldsymbol{f} = \widehat{\boldsymbol{f}}$ can be expressed by
\begin{equation*}
    P_{\boldsymbol{\eta}} \boldsymbol{f} := \Phi(\boldsymbol{\eta}) \Phi(\boldsymbol{\eta})^+ \boldsymbol{f},
\end{equation*}
where $\Phi(\boldsymbol{\eta}) := [\varphi_1(\boldsymbol{\eta}),\ldots,\varphi_m(\boldsymbol{\eta})]\in \mathbb{R}^{n \times m}$ and $\Phi(\boldsymbol{\eta})^+$ denotes its Moore-Penrose pseudoinverse. Because this projection depends on $\boldsymbol{\eta}$, $P_{\boldsymbol{\eta}}$ is often referred to as a variable projection (VP) operator. In addition to $P_{\boldsymbol{\eta}}$, we will make use of the feature transformation $C_{\boldsymbol{\eta}} : \mathbb{R}^n \to \mathbb{R}^m$ defined by
\begin{equation}
    \label{eq:coeffs}
    C_{\boldsymbol{\eta}}\boldsymbol{f} = \Phi(\boldsymbol{\eta})^+ \boldsymbol{f},
\end{equation}
which provides the coordinates of $\widehat{\boldsymbol{f}}$ with respect to the basis $\{\varphi_1(\boldsymbol{\eta}),\ldots,\varphi_m(\boldsymbol{\eta})\}$. This transformation can be interpreted as a dimension reduction and feature extraction step. If some a priori information is known about the behavior of $\boldsymbol{f}$ (e.g., quasi compactness, quasi periodicity, etc.) then the functions $\phi_{\boldsymbol{\eta}}^1,\ldots,\phi_{\boldsymbol{\eta}}^m$ can be chosen so that their parameter $\boldsymbol{\eta} \in \mathbb{R}^p$ contains physically meaningful information (see, e.g.,~\cite{weightedHermdozsa, dozsa2024variable, KOVACS}). Given such a family, it is reasonable to ask, if any $\boldsymbol{\eta}$ parameter minimizes 
\begin{equation}
    \label{eq:r2}
    \varrho(\boldsymbol{\eta}) := {\| \boldsymbol{f} - P_{\boldsymbol{\eta}}\boldsymbol{f} \|}_2^2.
\end{equation}
Minimizing the functional $\varrho$ in Eq.~\eqref{eq:r2} is also known as a SNLLS problem. In~\cite{varpro1}, Golub and Pereyra provide analytic formulas for the Jacobian of $\Phi(\boldsymbol{\eta)}^+$ and $P_{\boldsymbol{\eta}}$. Thanks to their contribution, minimizing Eq.~\eqref{eq:r2} can be done with gradient-based optimization schemes. This gives rise to a large number of applications (see, e.g.,~\cite{KOVACS} and~\cite{varpro2}).

\section{VP-SVM models}
Consider the dataset $\mathcal{X} := \left\{ \boldsymbol{f}_1,\ldots, \boldsymbol{f}_q\right\} \subset \mathbb{R}^n$ and its labeled version $\mathcal{D} := \left\{ (\boldsymbol{f}_1,y_1),\ldots, (\boldsymbol{f}_q,y_q)\right\} \subset \mathbb{R}^n \times\{-1,1\}$.

Our objective is to identify a nonlinear SVM $\mathcal{F}({\boldsymbol{\beta}};\cdot) : \mathbb{R}^n \to \{-1, 1\}$, $\mathcal{F}({\boldsymbol{\beta}};\boldsymbol{f}):= \textrm{sgn} \left( \sum_{j=1}^{q} \beta_j K(\boldsymbol{f}, \boldsymbol{f}_j) \right)$, 
%for which $(\boldsymbol{f}_j, \mathcal{F}_{\boldsymbol{\beta}}(\boldsymbol{f}_j)) \in \mathcal{D}$ for as many $j=1,\ldots,q$ as possible (which means a good training accuracy).
where $\boldsymbol{\beta} := (\beta_1,\ldots,\beta_q)^\top \in \mathbb{R}^q$ and $K(\cdot, \cdot)$ is a reproducing kernel of a reproducing kernel Hilbert space (RKHS)~\cite{stitson1996theory}. The purpose of $K$ is to lift the data samples $\boldsymbol{f}_j \in \mathcal{X}$, $j=1,\ldots,q$ to a higher dimensional space, where a separating hyperplane between the classes, defined by $\boldsymbol{\beta}$, might be found. In this work, following the reasoning in~\cite{dozsa2024variable}, we restrict our investigation to the  radial basis function (RBF) kernel defined by $K(\boldsymbol{f}, \boldsymbol{g}):= \mathrm{e}^{-\|\boldsymbol{f} - \boldsymbol{g}\|_2^2 / \sigma^2} \ (\boldsymbol{f}, \boldsymbol{g} \in \mathbb{R}^n)$ for some $\sigma \in \mathbb{R}\setminus\{ 0\}$. 
It is well-known~\cite{goodfellow2016deep} that an appropriately chosen  feature extraction transformation can significantly improve classification performance. In this work we will consider the transformation $ C_{\boldsymbol{\eta}}\boldsymbol{f}$, as defined in \eqref{eq:coeffs}, which is learned jointly with the weights of the underlying classifier, i.e.,
\begin{equation}
    \label{eq:vpsvmmodel}
    \begin{split}
        & \mathcal{F}(\boldsymbol{\beta}, \boldsymbol{\eta};\boldsymbol{f}) = \textrm{sgn} \left( \sum_{j=1}^{q} \beta_j K \left(C_{\boldsymbol{\eta}}\boldsymbol{f}, C_{\boldsymbol{\eta}}\boldsymbol{f}_j  \right) \right) \\ & (\boldsymbol{\beta} \in \mathbb{R}^q, \boldsymbol{\eta} \in \mathbb{R}^p, \boldsymbol{f} \in \mathbb{R}^n, \boldsymbol{f}_j \in \mathcal{X}).
    \end{split}
\end{equation}
The goal of training a VP-SVM classifier is to find appropriate parameters $\boldsymbol{\eta}$ and $\boldsymbol{\beta}$ for the model defined in Eq.~\eqref{eq:vpsvmmodel}. This can be done, for example (see~\cite{dozsa2024variable}), by solving
\begin{align}
    \label{eq:vpsvmobj}
    \begin{split}
        &\min_{\boldsymbol{\beta} \in \mathbb{R}^q, \boldsymbol{\eta} \in \mathbb{R}^p} e(\boldsymbol{\beta},  \boldsymbol{\eta})
        \\ &= \min_{\boldsymbol{\beta} \in \mathbb{R}^q, \boldsymbol{\eta} \in \mathbb{R}^p} \lambda_0 R(\boldsymbol{\eta}) + \lambda_1 A(\boldsymbol{\beta}, \boldsymbol{\eta}) + \lambda_2 B(\boldsymbol{\beta}, \boldsymbol{\eta}) , 
    \end{split}
\end{align}
where 
\begin{align*}
 R(\boldsymbol{\eta}) &:=\frac{1}{q} \sum_{j=1}^q \frac{{\|\boldsymbol{f}_j - P_{\boldsymbol{\eta}} \boldsymbol{f}_j\|}_2^2}{{\|\boldsymbol{f}_j\|}_2^2}, \\
 A(\boldsymbol{\beta},\boldsymbol{\eta}) &:= \boldsymbol{\beta}^\top K(\boldsymbol{\eta}) \boldsymbol{\beta}, \, K(\boldsymbol{\eta}) := \left[K \left(C_{\boldsymbol{\eta}}\boldsymbol{f}_i, C_{\boldsymbol{\eta}}\boldsymbol{f}_j  \right)\right]_{i,j=1}^q, \\
 B(\boldsymbol{\beta},\boldsymbol{\eta}) &:= \frac{1}{q}\sum_{j=1}^q \left( \max\big(0, 1 - y_j(\boldsymbol{\beta}^\top K(\boldsymbol{\eta})_j\big) \right)^2,
\end{align*}
and where $\lambda_0, \lambda_1, \lambda_2 > 0$ are weighting constants for the different terms, and 
$K_j$ denotes the $j$-th column of the matrix $K$. The quadratic loss term $B$ defines a classical SVM and is twice differentiable, except for the case where a sample is exactly on the separating hyperplane.
\section{second-order differentiation of VP-operators}
\label{sec:diffs}%
In our training approach, we will incorporate curvature information, we use first and second-order derivatives of the objective functional in Eq.~\eqref{eq:vpsvmobj}. While in practice automatic differentiation tools can produce these, we now proceed with providing the required formulas, focusing first on the regularization term $R(\boldsymbol{\eta})$ in Eq.~\eqref{eq:vpsvmobj}. The gradient of this term can be found in~\cite[p. 5, Eq. (9)]{KOVACS}, while its Hessian is given by
\begin{align}
    \label{eq:Rhess}
    \begin{split}
    & \frac{\partial^2 R(\boldsymbol{\eta})}{\partial \eta_j \partial \eta_i} = \frac{1}{q} \sum_{k=1}^{q} \frac{2}{\|\boldsymbol{f}_k\|^2_2} \left( \left\langle \frac{\partial P_{\boldsymbol{\eta}}\boldsymbol{f}_k}{\partial \eta_j}, \frac{\partial P_{\boldsymbol{\eta}}\boldsymbol{f}_k}{\partial \eta_i} \right\rangle -\right. \\ & \qquad \left. \left\langle \boldsymbol{f}_k - P_{\boldsymbol{\eta}}\boldsymbol{f}_k, \frac{\partial^2 P_{\boldsymbol{\eta}}\boldsymbol{f}_k}{\partial \eta_j \partial \eta_i} \right\rangle \right) \quad (i,j=1,\ldots,p).
    \end{split}
\end{align}
Notice that first- and second-order derivatives of the variable projection operator $P_{\boldsymbol{\eta}}$ are needed to evaluate Eq.~\eqref{eq:Rhess}. The first-order derivatives can be found in~\cite{dozsa2024variable}, while an elementary calculation shows that for $i,j=1,\ldots,p$,
\begin{align*}
  %  \label{eq:Phess}
    \begin{split}
        & \frac{\partial^2 P_{\boldsymbol{\eta}}}{\partial \eta_j\partial \eta_i}  = \ \frac{\partial P_{\boldsymbol{\eta}}^{\perp}}{\partial \eta_j} \frac{\partial \Phi(\boldsymbol{\eta})}{\partial \eta_i} \Phi(\boldsymbol{\eta})^+ + P_{\boldsymbol{\eta}}^{\perp} \frac{\partial^2 \Phi(\boldsymbol{\eta})}{\partial \eta_j\partial \eta_i} \Phi(\boldsymbol{\eta})^+ \\ & \qquad + P_{\boldsymbol{\eta}}^{\perp} \frac{\partial \Phi(\boldsymbol{\eta})}{\partial \eta_i} \frac{\partial \Phi(\boldsymbol{\eta})^+}{\partial \eta_j} + \left[ \frac{\partial P_{\boldsymbol{\eta}}^{\perp}}{\partial \eta_j} \frac{\partial \Phi(\boldsymbol{\eta})}{\partial \eta_i} \Phi(\boldsymbol{\eta})^+ \right. \\ & \qquad  \left. + P_{\boldsymbol{\eta}}^{\perp} \frac{\partial^2 \Phi(\boldsymbol{\eta})}{\partial \eta_j\partial \eta_i} \Phi(\boldsymbol{\eta})^+ + P_{\boldsymbol{\eta}}^{\perp} \frac{\partial \Phi(\boldsymbol{\eta})}{\partial \eta_i} \frac{\partial \Phi(\boldsymbol{\eta})^+}{\partial \eta_j}\right]^\top,
    \end{split}
\end{align*}
where $P_{\boldsymbol{\eta}}^{\perp} := I - P_{\boldsymbol{\eta}}$ denotes the orthogonal complement operator of $P_{\boldsymbol{\eta}}$ and its derivative is given by $\frac{\partial P_{\boldsymbol{\eta}}^{\perp}}{\partial \eta_j} = - \frac{\partial P_{\boldsymbol{\eta}}}{\partial \eta_j}$ accordingly. Finally, to fully express the Hessian of $R(\boldsymbol{\eta})$, we need to provide the second-order derivatives of the pseudoinverse $\Phi(\boldsymbol{\eta})^{+}$. Like before, the first-order formulas can be found in the literature (see,  e.g.,~\cite{chapelle2007training}), while the corresponding second partial derivatives for $i,j=1,\ldots,p$ are given by
\begin{align*}
    %\label{eq:Pinvhess}
    \begin{split}
    & \frac{\partial^2 \Phi(\boldsymbol{\eta})^+}{\partial \eta_j\partial \eta_i}  = -\frac{\partial \Phi(\boldsymbol{\eta})^+}{\partial \eta_j} \frac{\partial \Phi(\boldsymbol{\eta})}{\partial \eta_i} \Phi(\eta)^+ \\ & \qquad - \Phi(\boldsymbol{\eta})^+ \frac{\partial^2 \Phi(\boldsymbol{\eta})}{\partial \eta_j\partial \eta_i} \Phi(\boldsymbol{\eta})^+ - \Phi(\boldsymbol{\eta})^+ \frac{\partial \Phi(\boldsymbol{\eta})}{\partial \eta_i} \frac{\partial \Phi(\boldsymbol{\eta})^+}{\partial \eta_j} \\ & \qquad + \frac{\partial \Phi(\boldsymbol{\eta})^+}{\partial \eta_j} \left[\Phi(\boldsymbol{\eta})^+\right]^\top \frac{\partial \Phi(\boldsymbol{\eta})}{\partial \eta_i}^\top P_{\boldsymbol{\eta}}^{\perp} \\
    & \qquad + \Phi(\boldsymbol{\eta})^+ \left[\Phi(\boldsymbol{\eta})^+\right]^\top \frac{\partial \Phi(\boldsymbol{\eta})}{\partial \eta_i}^\top \frac{\partial P_{\boldsymbol{\eta}}^{\perp}}{\partial \eta_j} \\ & \qquad + \left( - \frac{\partial \Phi(\boldsymbol{\eta})^+}{\partial \eta_j} \Phi(\boldsymbol{\eta}) - \Phi(\boldsymbol{\eta})^+ \frac{\partial \Phi(\boldsymbol{\eta})}{\partial \eta_j} \right) \\ & \qquad \times \frac{\partial \Phi(\boldsymbol{\eta})^\top}{\partial \eta_i} \left[\Phi(\boldsymbol{\eta})^+\right]^\top \Phi(\boldsymbol{\eta})^+ \\
    & \qquad + \left(I - \Phi(\boldsymbol{\eta})^+ \Phi(\boldsymbol{\eta})\right) \frac{\partial^2 \Phi(\boldsymbol{\eta})^\top}{\partial \eta_j\partial \eta_i} \left[\Phi(\boldsymbol{\eta})^+\right]^\top \Phi(\boldsymbol{\eta})^+ \\
    & \qquad + \left(I - \Phi(\boldsymbol{\eta})^+ \Phi(\boldsymbol{\eta})\right) \frac{\partial \Phi(\boldsymbol{\eta})^\top}{\partial \eta_i} \frac{\partial \left[\Phi(\boldsymbol{\eta})^+\right]^\top}{\partial \eta_j} \Phi(\boldsymbol{\eta})^+ \\ & \qquad + \left(I - \Phi(\boldsymbol{\eta})^+ \Phi(\boldsymbol{\eta})\right) \frac{\partial \Phi(\boldsymbol{\eta})^\top}{\partial \eta_i} \left[\Phi(\boldsymbol{\eta})^+\right]^\top \frac{\partial \Phi(\boldsymbol{\eta})^+}{\partial \eta_j},
    \end{split}
\end{align*}
where $\frac{\partial^2 \Phi(\boldsymbol{\eta})}{\partial \eta_j\partial \eta_i}$, $i,j=1\dots,p$, are assumed to exist and be known for a given function system which defines $\Phi(\boldsymbol{\eta})$. 

The partial derivatives of $A$ and $B$ from Eq.~\eqref{eq:vpsvmobj} also need to be determined. For first-order and second-order derivatives with respect to $\boldsymbol{\beta}$, we refer to~\cite{chapelle2007training}. Assuming that the kernel $K(\boldsymbol{\eta})$ is twice differentiable, for $i=1,\ldots,p$ we obtain $\frac{\partial B(\boldsymbol{\beta},\boldsymbol{\eta})}{\partial \eta_i} = -\frac{2}{q} \sum_{\ell=1}^{q} \max\left( 0, 1- y_\ell \boldsymbol{\beta}^\top {K(\boldsymbol{\eta})}_\ell\right) y_\ell \frac{\partial K(\boldsymbol{\eta})}{\partial \eta_i } \boldsymbol{\beta}$.
%\begin{equation*}
   % \label{eq:gradObjB}
    %\frac{\partial B(\boldsymbol{\beta},\boldsymbol{\eta})}{\partial \eta_i} = -\frac{2}{q} \sum_{\ell=1}^{q} \max\left( 0, 1- y_\ell \boldsymbol{\beta}^\top {K(\boldsymbol{\eta})}_\ell\right) y_\ell \frac{\partial K(\boldsymbol{\eta})}{\partial \eta_i } \boldsymbol{\beta}.
%\end{equation*}
 The second-order partial derivatives of $B(\boldsymbol{\beta}, \boldsymbol{\eta})$ for $i,j=1,\ldots,p$ are then given by
\begin{align*}
%\label{eq:hessObjB}
\begin{split}&\frac{\partial^2 B(\boldsymbol{\beta},\boldsymbol{\eta})}{\partial \eta_j\partial \eta_i} = -\frac{2}{q} \sum_{\ell=1}^{q} \left[-\max\left(0, 1- y_\ell \boldsymbol{\beta}^\top \frac{\partial {K(\boldsymbol{\eta})}_\ell}{\partial \eta_j}\right) y_\ell \right. \\
& \,  \times  \left. \frac{\partial K(\boldsymbol{\eta})}{\partial \eta_i } \boldsymbol{\beta} +  \max\left(0, 1- y_\ell
\boldsymbol{\beta}^\top {K(\boldsymbol{\eta})}_\ell\right) y_\ell \frac{\partial^2 K(\boldsymbol{\eta})}{\partial\eta_j\partial \eta_i } \boldsymbol{\beta} \right].
\end{split}
\end{align*}
 Furthermore, for $i,j=1,\ldots,p$ we obtain
\begin{equation*}
    %\label{eq:gradhessA}
    \frac{\partial A(\boldsymbol{\beta}, \boldsymbol{\eta})}{\partial \eta_i}  = \boldsymbol{\beta}^\top \frac{\partial K(\boldsymbol{\eta})}{\partial \eta_i} \boldsymbol{\beta}, \quad \frac{\partial^2 A(\boldsymbol{\beta}, \boldsymbol{\eta})}{\partial \eta_j \partial \eta_i} = \boldsymbol{\beta}^\top \frac{\partial^2 K(\boldsymbol{\eta})}{\partial \eta_j \partial \eta_i} \boldsymbol{\beta}.
\end{equation*}
Finally, the mixed derivatives are expressed by
\begin{equation}
    \label{eq:gradMixed}
    \begin{split}
       & \frac{\partial^2 e(\boldsymbol{\beta}, \boldsymbol{\eta})}{\partial \boldsymbol{\eta} \partial \boldsymbol{\beta}} = 2\lambda_1 \frac{\partial K(\boldsymbol{\eta})}{\partial \boldsymbol{\eta}} \boldsymbol{\beta}  \\ & \qquad + \frac{2\lambda_2}{q}  \left( \frac{\partial K(\boldsymbol{\eta})}{\partial \boldsymbol{\eta}} \Pi (K(\boldsymbol{\eta})\boldsymbol{\beta} - \boldsymbol{y}) + K(\boldsymbol{\eta}) \Pi \frac{\partial K(\boldsymbol{\eta})}{\partial \boldsymbol{\eta}} \boldsymbol{\beta} \right).
    \end{split}
\end{equation}
In Eq.~\eqref{eq:gradMixed}, we assume that the first $r \in \mathbb{N}$ entries of the label vector $\boldsymbol{y} = (y_1,\ldots,y_q)^\top$ are exactly the ones for which $1 - y_i \boldsymbol{\beta}^\top K(\boldsymbol{\eta})_i > 0$. Furthermore, we define the diagonal matrix $\Pi \in \mathbb{R}^{q \times q}$ by $\Pi := \operatorname{diag}(\pi_1,\ldots,\pi_q)$, where \vspace{-2mm}
\begin{equation*}
    \pi_k = \begin{cases}
        1, & k \leq r \\
        0, & \textrm{otherwise}
    \end{cases}, \quad k=1,\ldots,q.
\end{equation*}

\section{Experiments}
\label{sec:exp}

We now show how second-order information can be used to train a VP-SVM classifier to recognize road surface abnormalities. We rely on the dataset introduced in~\cite{roadbnorm}. The dataset contains a total of $q=517$ 1D signals recorded on the public road network of Budapest. The signals correspond to voltage changes in a tire-implanted force sensor and thus to tire deformation. The signals have been segmented into single tire revolutions and zero-padded, producing compactly supported waveforms of equal length. %The support of these signals changes with vehicle velocity and the waveforms are characterized by a main lobe that indicates the point of largest deformation (when the sensor was closest to the surface). 
For information about measurements and signal characteristics, we refer to~\cite{roadbnorm}.

The first few  adaptive Hermite functions~\cite{ecgDozsa} can be used to effectively approximate the tire signals~\cite{roadbnorm}. For this reason, we choose $\varphi_k(\boldsymbol{\eta})$, $k=1,\ldots,m$ as the discrete, equidistant sampling of the $k$-th adaptive Hermite function~\cite{weightedHermdozsa} defined by
\begin{multline*}
    \label{eq:hsys}
  \varphi_k(\boldsymbol{\eta}) =  \left[\sqrt{\frac{s}{\pi^{1/2}2^k k!}} h^k(s(x_\ell - t)) \mathrm{e}^{-(s(x_\ell-t))^2/2}\right]_{\ell=1}^n  \\ 
         (s > 0, \ x_\ell, t \in \mathbb{R}, \ \boldsymbol{\eta} := (s, t)^\top \in \mathbb{R}^2),
\end{multline*}
where $h^k$ denotes the $k$-th classical Hermite polynomial~\cite{szeg1939orthogonal}. %For valid choices of $\boldsymbol{\eta} = (s,t)^\top$, adaptive Hermite functions form a complete and orthonormal system in $L_2(\mathbb{R})$ and have been successfully applied in various VP-based machine learning methods~\cite{KOVACS, dozsa2024variable, weightedHermdozsa}. 
Importantly, for adaptive Hermite functions, the existence of a global solution for minimizing Eq.~\eqref{eq:r2} is proven in~\cite{weightedHermdozsa}. Finally, we note that the dilation parameter $s$ can be used to model the support of the signals and thus vehicle speed, making the representation interpretable. In~\cite{roadbnorm}, a total of $11$ adaptive Hermite functions are proposed to obtain the features. VP-based neural networks then achieve an average of $97 \%$ accuracy on the dataset. In contrast, we use the much simpler VP-SVM model and represent the signals using only $5$ Hermite-Fourier coefficients. Regardless, as seen in Table~\ref{tab:accuracy}, our proposed training scheme is able to achieve performance close to the state-of-the-art.

%%%%%%%%%%%%%
%The convexity of the objective $e$ from Eq.~\eqref{eq:vpsvmobj} cannot be assumed due to the introduction of the variable projection operator (see also examples of negative Hessians in our implementation~\cite{code}). In order to mitigate this, we propose the use of trust-region (TR) algorithms to solve~\eqref{eq:vpsvmobj}.

In particular, we compare the performance of gradient descent (GD) and the TR method proposed in Algorithm 6.1.1 of \cite{CONN} for training VP-SVM (see Eq.~\eqref{eq:vpsvmobj}). The GD method uses first-order information only. Note that performing a line search along the negative gradient direction would be prohibitive due to the multiple kernel evaluations required at each iteration. Therefore, we fix a small learning rate of $10^{-3}$, as commonly done in stochastic gradient methods. In contrast, the TR approach leverages the second-order derivative discussed earlier to define the model and determine the search direction as an approximate solution to the trust region subproblem (using the Steihaug-Toint method, as outlined in \cite{STmethod}).
To the best of our knowledge, this is the first time second-order methods have been used to train a kernel-based classifier, where the kernel also depends on learnable parameters.
A key advantage of TR methods is that when a step is rejected, re-evaluating the Hessian along with 1st and 2nd kernel derivative evaluations is not required. 
Another advantage over approaches that rely on multiple objective function evaluations is that, in the case of a rejected step, TR only requires a single kernel evaluation per iteration.

All experiments are performed with initial weights $\beta=0$ and  parameters $\boldsymbol{\eta} = (s,t)^\top = (10,0)^\top$ for dilation and translation, fixed regularization parameters $\lambda_0 = 2$, $\lambda_1 = 0.1$, and $\lambda_2 = 1$.\footnote{All experiments were conducted on a workstation with an  Intel(R) Xeon(R) W-2123 CPU @ 3.60GHz, 64 GB RAM, and Python 3.10.12 with Pytorch 2.4.0 using automatic differentiation. All models were trained using the  CPU.}
\begin{figure}[h]
    \centering \resizebox{0.935\columnwidth}{!}{\input{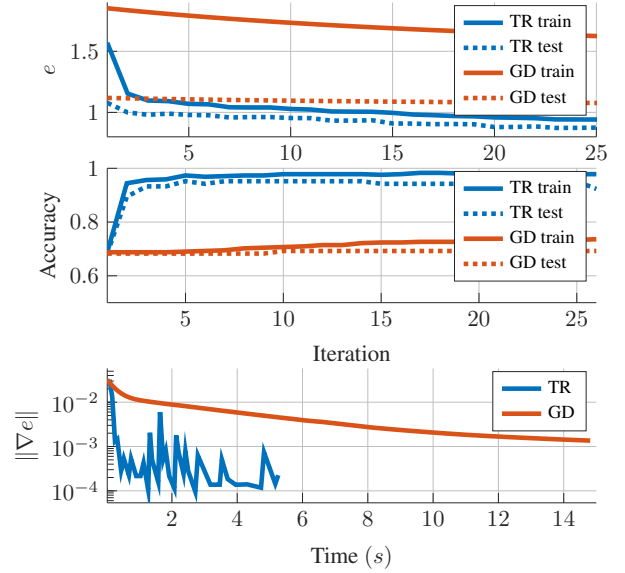}}%TikZ/triple_figure_VERTICAL_NEW_25iter.tikz
    \caption{Top and middle: Objective function and accuracy for both TR and GD on training and test sets.
    Bottom: Gradient norm evolution over time for TR and GD on the training set.}
    \label{fig:comparison}
\end{figure}
From Figure~\ref{fig:comparison}, we observe that the TR method achieves a faster reduction in the loss and higher accuracy compared to GD on both training and test sets. The bottom plot confirms that the TR method effectively reduces the gradient norm, whereas GD exhibits slower decay. %and more oscillatory decay.
% \begin{table}[h]
%     \centering
%         \caption{Summary of VP-SVM classification performance.
%     }
%     \label{tab:accuracy}
% \begin{tabular}{l@{\hskip 1pt}r@{\hskip 4pt}r@{\hskip 4pt}r@{\hskip 4pt}r@{\hskip 4pt}r@{\hskip 4pt}r}
%     \toprule
%     Alg. & $\#$It & Avg Acc. & Min Acc. & Max Acc. & $\|\nabla e\|$ & Time (s) \\
%     \midrule
%     TR & 15   & 94.0\% & 92.3\% & 96.2\% & 2.03 & $\boldsymbol{1.32}$ \\
%        & 25   & $\boldsymbol{95.7\%}$ & $\boldsymbol{94.2\%}$ & $\boldsymbol{97.1\%}$ & $\boldsymbol{0.03}$ & 6.72 \\ 
%     \midrule
%     GD & 25   & 85.9\% & 82.7\% & 90.3\% & 11.11 & $\boldsymbol{0.49}$ \\
%        & 125  & 86.5\% & 84.6\% & 89.3\% & 6.66 & 2.46 \\
%        & 250  & 86.9\% & 83.5\% & $\boldsymbol{91.3\%}$ & 6.88 & 4.90 \\
%        & 500  & 87.2\% & 84.5\% & 90.3\% & 2.02 & 10.09 \\
%        & 1000 & $\boldsymbol{88.4\%}$ & $\boldsymbol{85.4\%}$ & 90.3\% & $\boldsymbol{0.59}$ & 20.05 \\
%     \bottomrule
% \end{tabular}
% \end{table}
\begin{table}[h]
    \centering
        \caption{Summary of VP-SVM classification performance.
    }
    \label{tab:accuracy}
\begin{tabular}{l@{\hskip 0.3pt}r@{\hskip 2pt}r@{\hskip 2pt}r@{\hskip 2pt}r@{\hskip 3pt}r@{\hskip 3pt}r}
    \toprule
    Alg. & $\#$It & Avg Acc. & Min Acc. & Max Acc. & $\|\nabla e\|$ & Time (s) \\
    \midrule
    TR & 15   & 95.5\% & 92.2\% & $\boldsymbol{98.1\%}$ & 9.16e-04 & $\boldsymbol{1.73}$ \\
       & 25   & $\boldsymbol{96.3\%}$ & $\boldsymbol{93.2\%}$ & $\boldsymbol{98.1\%}$ & $\boldsymbol{2.24e}$-$\boldsymbol{04}$ & 5.39 \\ 
    \midrule
    GD & 25   & 72.3\% & 68.0\% & 77.9\% & 1.57e-02 & $\boldsymbol{0.33}$ \\
       & 125  & 83.9\% & 79.6\% & 88.5\% & 7.52e-03 & 1.68 \\
       & 250  & 88.2\% & 85.4\% & 93.3\% & 4.82e-03 & 3.39 \\
       & 500  & 91.3\% & 89.3\% & 95.2\% & 2.10e-03 & 6.78 \\
       & 1000 & $\boldsymbol{92.1\%}$ & $\boldsymbol{90.3\%}$ & $\boldsymbol{96.2\%}$ & $\boldsymbol{8.24e}$-$\boldsymbol{04}$ & 13.63 \\
    \bottomrule
\end{tabular}
\end{table}

Table~\ref{tab:accuracy} reports the classification performance obtained by training VP-SVM. Both the GD and TR solvers are evaluated using a $5$-fold cross-validation. The table summarizes the mean gradient norms $\|\nabla e\|$ and accuracies, together with the minimum and maximum values across the different tests.
The results show that TR consistently outperforms GD, achieving higher accuracy with substantially fewer iterations and time. TR reaches a given performance level much faster, whereas GD requires more than two times the training time and still fails to achieve TR’s accuracy. %This efficiency gap arises primarily from the repeated kernel evaluations in GD, which make each iteration costly. By contrast, TR requires fewer but more effective iterations, thereby reducing this overhead and delivering superior performance.

\section{Conclusion}
\label{sec:conc}

In this paper, we proposed second-order training schemes for VP-SVM classifiers. We provided the necessary analytic formulas and demonstrated the effectiveness of TR training schemes using second-order information.% for the second derivatives of variable projection operators and SVM. %objectives. In addition, we demonstrated the effectiveness of TR training schemes using second-order information for VP-SVM. %We verified the efficiency of the proposed methods through numerical experiments and provided publicly available codes~\cite{code} which can be used to reproduce our results. In the next phase of this research, we plan to combine our training schemes with well-known kernel reduction techniques (such as~\cite{fine2001efficient}) and extend our results to second-order stochastic optimization techniques.

{\begin{center}
{\bf DATA AND CODE AVAILABILITY\vspace{-.5em}\vspace{0pt}}
\end{center}}
Data and code to reproduce the results are available in~\cite{code}.%The data and code for reproducing the numerical results presented in this paper are available in \cite{code}.

{\begin{center}
{\bf ACKNOWLEDGMENT\vspace{-.5em}\vspace{0pt}}
\end{center}}
The research of TD has received funding from the Swiss Govern\-ment Excellence Scholarship No. 2025.0057. The research of AA was funded in part by the Swiss National Science Foundation (SNSF) grant No. 224943. The project was supported by the University Excellence Fund of E\"otv\"os Lor\'and University, Budapest, Hungary (ELTE). Project no. K146721 has been implemented with the support provided by the Ministry of Culture and Innovation of Hungary from the National Research, Development and Innovation Fund, financed under the K\_23 "OTKA" funding schemes.

\bibliographystyle{IEEEbib}
\bibliography{refs}
%\bibliography{strings,refs}

\end{document}